\documentclass[aps,a4paper,twocolumn,floatfix,showkeys,
groupaddress]{revtex4}

\usepackage{bm,amsmath,amssymb,amsfonts}

\usepackage[dvips]{graphicx}
\usepackage[colorlinks=true,citecolor=magenta,urlcolor=red,filecolor=blue]{hyperref}
\usepackage{color}
\definecolor{dgreen}{rgb}{0.0,0.5,0.0}
\begin{document}
\title{\textcolor{blue}
{Controlled trapping of single particle states on a periodic substrate 
by deterministic stubbing}}

\author{Amrita Mukherjee}
\email{amritaphy92@gmail.com}
\affiliation{Department of Physics, University of Kalyani, Kalyani,
West Bengal-741 235, India}
\author{Atanu Nandy}
\email{atanunandy1989@gmail.com}
\affiliation{Department of Physics, Kulti College, Kulti, Paschim Bardhaman, West Bengal - 713343, India}
\author{Arunava Chakrabarti}
\email{arunava_chakrabarti@yahoo.co.in}
\affiliation{Department of Physics, University of Kalyani, Kalyani,
West Bengal-741 235, India}






\begin{abstract}

A periodic array of atomic sites, described within a tight binding formalism 
is shown to be capable of trapping electronic states as it grows in size and 
gets stubbed by an `atom' or an `atomic' clusters from a side in a deterministic way.
We prescribe a method based on a real space renormalization group method, that 
unravels a subtle correlation between the positions of the side coupled atoms 
and the energy eigenvalues for which the incoming particle finally gets trapped.
We discuss how, in such conditions, the periodic backbone gets transformed into 
an array of infinite quantum wells in the thermodynamic limit. We present a case here, 
where the wells have a hierarchically distribution of widths, hosing standing wave solutions 
in the thermodynamic limit. 

\end{abstract}



\maketitle

\section{Introduction}
\label{sec1}
The ubiquitous and unavoidable phenomenon of Anderson localization 
in disordered systems has been alive for almost six 
decades now~\cite{anderson,abrahams,kramer}, with variations and surprises 
that have gone well beyond the realm 
of electronic systems, and incorporate phonons~\cite{monthus}, 
magnons~\cite{lyo}, photonics~\cite{john, yablono} and matter waves 
~\cite{roati} in recent times. The pivotal result in this field is that, 
all the single particle 
eigenfunctions remain exponentially localized in one and in two dimensions, irrespective 
of the strength of disorder. In three dimensions however, there remains 
a possibility of observing 
mobility edges separating the localized states from the extended, current carrying ones.

With the emergence of nano-fabrication techniques, interesting variants of Anderson 
localization can be put to test. One non-trivial variation of incorporating disorder is 
by coupling, from a side, a group of atoms that functionalize an otherwise periodically 
ordered backbone, which might mimic a quantum wire~\cite{orellana,grosso}. 
The coupling of discrete structures 
from a side initiates the incorporation of bound states in the continuum, giving rise to 
the interesting Fano line shapes in the transmission spectrum~\cite{mirosh,sergej}, 
and even in the density 
of states of the backbone-adatom system. 

In spite of the existing canonical cases of 
disorder induced localization in one dimensional systems, 
short range positional correlation has been shown to lead to unscattered, extended 
single particle states for special values of electron energies~\cite{dunlap}. The 
positional correlations among the constituents were exploited for quasiperiodically 
ordered one dimensional chains, leading to the observation of an infinite number of 
such unscattered eigenstates, obtained by exploiting the inherent self similarity of 
the concerned lattices~\cite{ac1,ac2}. 
Recently, the minimal quasi-one dimensionality introduced by the side coupled atomic clusters, 
in various topological shapes, have been shown to lead even to a complete delocalization 
of the electronic states, over the entire (or, almost the entire) range of the 
permissible spectrum~\cite{atanu, biplab1, biplab2}.   
\begin{figure}[ht]
{\centering \resizebox*{8cm}{3cm}{\includegraphics{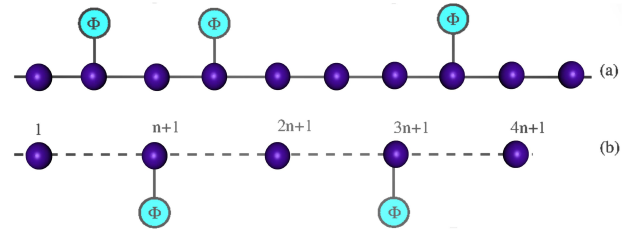}}\par}
\caption{(Color online). (a) A periodic chain of atomic sites, acting as the 
`backbone' is stubbed by single level quantum dots at positions 
following a middle third Cantor sequence. (b) The dots are now 
side coupled to a renormalized version of the same lattice that has been obtained 
by decimating out a set of $n$ sites uniformly.}
\label{lattice1}
\end{figure}

In this communication, our focus is on a different aspect. We consider a 
periodically ordered linear lattice, where all the single particle states are Bloch 
functions, extended and transparent, over a range of energy $[\epsilon-2t,\epsilon+2t]$,
when we use a tight binding Hamiltonian with nearest neighbor hopping 
approximation to describe the lattice. 
In the standard language, $\epsilon$ and $t$ are referred to as the `on site' 
potential and the nearest neighbor `hopping integral' respectively.
This 
lattice is then 'stubbed' deterministically, at predictably selected  
set of sites,  infinite in number, when 
the periodic backbone is also infinite in size. The side coupled 
entities can be 
single level `quantum dot's (QD) or quantum rings threaded by 
a magnetic flux. Both the `dot' and the `ring' are modelled in the same 
tight binding formalism. 

We show that, for a specified, completely deterministic set of 
energy eigenvalues, an electron travelling in such a 
stubbed lattice will eventually get trapped 
in quantum wells, with a hierarchical 
distribution of widths. The values of the {\it trapping energy} 
can be engineered 
at will, covering for example, the entire range for which the periodic backbone 
supported extended, transparent Bloch states. This can be accomplished by stubbing 
the backbone from a side with the QD's (or rings) at 
vertices farther and farther apart in the parent chain. The locations of such vertices
can be easily determined via a real space renormalization group (RSRG) 
decimation technique. With
increasing inter-stub separations, the distribution of the `trapping energy' eigenvalues 
gets more and more densely populated, being driven towards a continuous distribution 
in the thermodynamic limit. Even with a finite value of the potential offered by the 
defected sites, the lattice, in the thermodynamic limit, 
 turns out to be consisting of an infinite array of quantum 
wells with the heights of the walls growing finally to infinity, 
following a power law, as revealed {\it only} by 
repeated length scaling of the initial stubbed backbone.


In what follows we describe the method and the central results. In section I we
describe the Hamiltonian and the basic scheme of the work, describing in details
the RSRG scheme. Section II discusses the one dimensional periodic chain stubbed 
from a side with geometric structures. Two cases 
are discussed. First, we explain the way to trap electrons over 
the entire range of eigenvalues of a linear periodic chain. This is done 
using a single level QD. In the second example, we show 
how to control the energy values at
will by fixing quantum rings threaded by a magnetic flux at 
a set of sites 
throughout the 
infinite backbone. Section III discusses the results, and in 
section IV we discuss the basic principle of trapping 
light waves using an elementary model for monomode wave guides which 
is stubbed at suitable points.  
In section V we draw conclusions. 

\textbf{The model and the preliminaries.} -- 
We describe our system of a linear periodic lattice in the tight binding 
formalism, using the Hamiltonian for spinless, non-interacting electrons with 
nearest neighbor interaction:
\begin{equation}
H = \epsilon \sum_{j} |j\rangle \langle j| + t \sum_{<jk>} (|j\rangle \langle k| + h.c.)
\label{ham}
\end{equation}
where, $\epsilon$ and $t$ are the constant on-site potential and uniform nearest neighbor 
hopping integral respectively. The index $k$ thus has values $j \pm 1$.
The Schr\"{o}dinger equation for this lattice is written in an equivalent form of a 
difference equation for every site $j$ as,  
\begin{equation}
(E - \epsilon) \psi_j = t (\psi_{j-1} + \psi_{j+1})
\label{diff1d}
\end{equation}

We now digress a bit to remind ourselves 
 an interesting property of a one dimensional fractal, viz, 
a middle third Cantor sequence.
The Cantor array is sequentially built up using two characters, say $L$ and $S$. 
We take them to represent two kinds of `bonds' in a linear atomic chain. The growth 
rule of the Cantor sequence we discuss here is, $L \rightarrow LSL$ and 
$S\rightarrow SSS$, beginning with an $L$ bond (the first generation). 
The first three generations will read, $G_1=L$, 
$G_2=LSLSSSLSL$ and $G_3=LSLS^3LSLS^9LSLS^3LSL$, and one can 
figure out the shape of the lattice at any $n$-th stage, $G_n$. 
A tight binding description of electronic states 
using the Hamiltonian in Eq.~\eqref{ham} for such a lattice 
requires the identification 
of three different kinds of on-site potentials, namely, $\epsilon_\alpha$, 
$\epsilon_\beta$ and $\epsilon_\gamma$, for the atomic sites residing between 
the pair of bonds $L-L$, $L-S$ and $S-L$ respectively. The nearest neighbor 
hopping integrals are named $t_L$ or $t_S$ depending on the bond $L$ or $S$ 
the electron hops across.
\begin{figure}[ht]
{\centering \resizebox*{8cm}{2cm}{\includegraphics{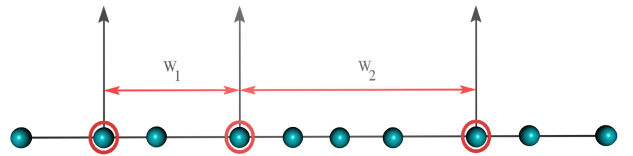}}\par}
\caption{(Color online). Schematic representation of an $n$ times renormalized 
Cantor lattice with $E=\epsilon_\alpha=\epsilon_\gamma \ne \epsilon_\beta$ and 
$t_L=t_S$. The renormalized $\beta$ sites (encircled in red) erect infinitely 
large effective value of the on-site potential as $n \rightarrow \infty$. The 
widths $W_1$, $W_2$ follow a hierarchical distribution, dictated by the Cantor 
sequence.}  
\label{well}
\end{figure}

Because of the inherent self similar hierarchical way 
the lattice builds up, it is easy to renormalize any $n$-th generation back to the 
$n-1$th generation by reversing the inflation rule. The process of renormalization
leads to the recursion relations for the potential and the hopping terms, 
that are given by~\cite{sheelan},
\begin{eqnarray}
\epsilon_{\alpha,n} & = & \epsilon_{\alpha,n-1} + \frac{2 t_{S,n-1}^2 
(E-\epsilon_{\alpha,n-1})}{\Delta_{1,n-1}} \nonumber \\
\epsilon_{\beta,n} & = & \epsilon_{\beta,n-1} + \frac{t_{L,n-1}^2 (E-\epsilon_{\beta,n-1})}
{\Delta_{2,n-1}} + \frac{t_{S,n-1}^2 (E-\epsilon_{\alpha,n-1})}{\Delta_{1,n-1}} \nonumber\\
\epsilon_{\gamma,n} & = & \epsilon_{\gamma,n-1} + \frac{t_{L,n-1}^2 (E-\epsilon_{\gamma,n-1})}
{\Delta_{2,n-1}} + \frac{t_{S,n-1}^2 (E-\epsilon_{\alpha,n-1})}{\Delta_{1,n-1}} \nonumber\\
t_{L,n} & = & \frac{t_{L,n-1}^2 t_{S,n-1}}{\Delta_{2,n-1}} \nonumber \\
t_{S,n} & = & \frac{t_{S,n-1}^3}{\Delta_{1,n-1}}
\label{recursion}
\end{eqnarray}
where, $\Delta_{1,n}=(E-\epsilon_{\alpha,n})^2-t_{S,n}^2$ and, 
$\Delta_{2,n}=(E-\epsilon_{\beta,n})(E-\epsilon_{\gamma,n})-t_{S,n}^2$. 

An interesting observation can be made using the set of Eq.~\eqref{recursion}. If we 
set $E = \epsilon_\alpha = \epsilon_\gamma \ne \epsilon_\beta$, and $t_L = t_S$ at the 
bare length scale, then with successive renormalization, the parameters evolve as, 
$\epsilon_{\alpha,n+1} = \epsilon_{\gamma,n+1}$, 
$\epsilon_{\beta,n+1} = 2 \epsilon_{\beta,n} 
-2$, and $t_{L,n+1}=t_{S,n+1}=-t_{L,n}=-t_{S,n}$ for $n \ge 0$. Two important aspects 
are to be noted here. First, we observe that the {\it effective} on-site 
potential at the $\beta$ sites grows with every iteration, and second, the 
hopping integrals remain equal in magnitude 
at every stage of RSRG operation, and get locked 
into a two cycle fixed point. This implies that, in the limit $n \rightarrow \infty$, 
which essentially means that we are looking at the lattice 
in its thermodynamic limit, the incoming electron (with energy $E = \epsilon_\alpha$)
will see an array of infinite potential wells of a {\it hierarchy of widths}.
Following Lindquist and Riklund~\cite{lind}, 
it can be understood that, in the limit of infinite 
well depths (achieved only when the lattice extends to infinity), standing  
waves are formed in the quantum wells trapping the electron. 
As observed in the literature~\cite{lind}, the envelope of the standing well 
in a quantum well comprising $N$ discrete sites is given 
by, 
\begin{equation}
\psi_{m,n} = \sqrt{\frac{2}{N+1}} \sin \left (\frac{mn\pi}{N+1} \right)
\label{stand1}
\end{equation}

We exploit this phenomenon to trap an electron travelling in linear
periodic lattice, as explained below.

\textbf{Trapping an electron on a linear ordered chain.} -- 
\textbf{Blocking propagation over the energy range $[\epsilon-2t,\epsilon+2t]$.} --

Let us refer to Fig.1. The periodic chain is depicted with black circles 
at the atomic sites. We `artificially' assign the bonds the `names' $L$ and $S$ from the left 
following the growth rule of the Cantor sequence as discussed in the last section.
Then it is easy to identify the $\beta$ sites, as well as the 
$\alpha$ and the $\gamma$ ones. We must appreciate that this design is artificial and the 
`pure' character of the host lattice doesn't 
change in terms of its energy spectrum and the 
nature of the eigenfunctions. Next we attach, in the simplest case, 
an atom, mimicking a single level QD from one side (cyan circles) to each `$\beta$' site.
The QD, assigned the same on-site potential $\epsilon$ as the 
backbone, is coupled to the backbone through a tunnel hopping $\lambda$.
This immediately makes the lattice resemble a Cantor chain in the site model, where 
we automatically ensure $\epsilon_\alpha=\epsilon_\gamma \ne \epsilon_\beta$ and 
$t_L = t_S$ at the bare length scale. We then iterate the recursion 
relations Eq.~\eqref{recursion} with the initial values, 
$\epsilon_\alpha=\epsilon_\gamma=
\epsilon$, $\epsilon_\beta = \lambda^2/(E-\epsilon)$, which is 
obtained by renormalizing the self energy at the $\beta$ site (by `folding' back the 
adatom into it), and $t_L=t_S=t$.

We have assigned a constant value 
of the on-site potential, viz. $\epsilon$ to the backbone, and the 
same potential is assigned to the attached QD as well. Next, the energy is 
chosen to be $E = \epsilon_\alpha = \epsilon_\gamma = \epsilon$. The recursion 
relations evolve and with every iteration the effective, renormalized potential 
at the $\beta$ site increases following the rule $\epsilon_{\beta,n} = 2 \epsilon_{\beta,n-1} -2$, 
while the renormalized nearest neighbor hopping integrals get locked into a two cycle fixed 
point, implying that the wave functions at two nearest neighbors at any scale of length 
has non-zero overlap. This implies that the states, in the limit $n \rightarrow \infty$, are 
standing waves, trapping the electron with energy $E = \epsilon$ eventually, in the thermodynamic limit.
The energy $E=\epsilon$ is the center of the band of the ordered substrate, without stubs.  
 
The scheme outlined above gives us an opportunity to engineer the energy values 
at which we can trap the propagating electron at will. To achieve this goal, we 
renormalize the parent lattice (without adatoms) first, by 
sequentially decimating clusters of 
$n$ atoms. This leads to the renormalization of the on-site potential 
and the nearest neighbor 
hopping integrals to,
\begin{eqnarray}
\epsilon_{eff}(E) & = & \epsilon + \frac{2t U_{n-1}(x)}{U_n(x)} \nonumber \\
t_{eff}(E) & = & \frac{t}{U_{n}(x)}
\label{rgeps}
\end{eqnarray}
It is now necessary to identify the {\it artificial} $\beta$ sites on this 
renormalized lattice and attach the single level QD 
to them, to make the renormalized system resemble again a Cantor sequence of sites, at a larger 
scale of length and with energy dependent on-site potentials and hopping integrals. If we 
map back to the original scale of length then it is obvious that now the stubs are places 
farther apart.
The final effective potential at the $\beta$ site on this rescaled version of the backbone is, 
$\epsilon_{\beta}^{eff} = \epsilon_{eff}(E) + \lambda^2/(E-\epsilon)$. The set of parameters with which 
the recursion relations Eq.~\eqref{recursion} now begin to iterate is, 
$\epsilon_{\alpha,0} = \epsilon_{\gamma,0} = \epsilon+2t U_{n-1}(x)/U_{n}(x)$, 
$\epsilon_{\beta,0} = \epsilon+2t U_{n-1}(x)/U_{n}(x)+\lambda^2/(E-\epsilon)$, 
$t_{L,0}=t_{S,0}=t/U_{n}(x)$. Here, $x=(E-\epsilon)/2t$.

We now need to look for the solutions of the equation $E - \epsilon_{eff}(E) = 0$, which 
is a polynomial in $E$. The real roots of this equation make the potential at the 
$\beta$ sites grow , and the electron faces unsurmountable barrier in the 
thermodynamic limit and eventually gets trapped in quantum wells , but now of much larger 
widths (that is, $L$ is larger now). To quote certain specific values, we can 
mention that with $n=1$, the roots of the polynomial equation are $E=\epsilon \pm 2t$.
With $\epsilon=0$ and $t=1$, the values are $\pm \sqrt{2}$. For $n=2$, $3$ and $4$, the 
`trapping' energy values are, $E=(0, \pm \sqrt{3})$, $(\pm 0.76537, \pm 1.84776)$, and 
$(0, \pm 1.17557, \pm 1.90211)$ respectively. Extending the idea we can understand that, 
by placing the adatoms wider and wider apart in the original lattice, which amounts to 
fixing them at the `artificial' $\beta$ sites at higher scales of renormalization (achieved
by increasing the value of $n$), we can in principle, obtain arbitrarily large 
number of the trapping energies which will ultimately densely fill the range of allowed 
energy eigenvalues of the periodic backbone, viz., $[\epsilon-2t,\epsilon+2t]$.

\begin{figure}[ht]
{\centering \resizebox*{7cm}{3cm}{\includegraphics{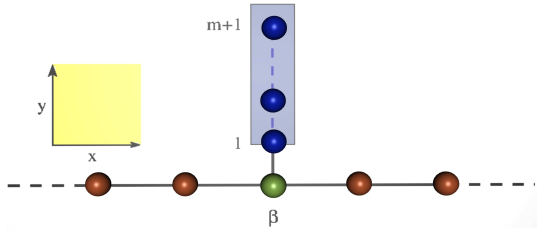}}\par}
\caption{(Color online). Schematic representation of an $n$ times renormalized 
periodic lattice with  
an $m+1$ atom cluster attached from one side on an infinite subset 
of sites so that such sites get the status of $\beta$ sites (in green) 
in an `effective' Cantor sequence.} 
\label{nontrap}
\end{figure}

\textbf{Does it always trap?} -- 
We discuss an obvious issue. Does a stub of {\it any length} 
can be used to trap an electron? To address this issue, We refer to Fig.~\ref{nontrap}. 
A periodically ordered chain of atomic sites is renormalized by 
decimating clusters of $n$ atoms sequentially. The renormalized values of the 
on-site potential and the hopping integrals in this lattice are given by Eq.~\eqref{rgeps}.
This `renormalized' lattice, exactly in the spirit of the earlier discussion, is 
`converted' into a Cantor sequence of sites by attaching an $m+1$ atom stub at every 
$\beta$-site, colored red in Fig.~\ref{nontrap}.
The effective $\beta$-site of the {\it converted} Cantor chain is now
given by, 
\begin{equation}
\epsilon_\beta = \epsilon+2t \frac{U_{n-1}(x)}{U_n(x)} + 
\frac{\lambda^2 U_m(y)}{(E-\epsilon) U_m(y)-\lambda U_{m-1}(y)}
\label{notrap}
\end{equation} 
Here, as before, $x=(E-\epsilon)/2t$, and $y=(E-\epsilon)/2\lambda$, where, 
$\lambda$ is the nearest neighbor hopping integral along the cluster in the 
$Y$-direction and has a non-zero value.
 Needless to say that, along the backbone that is obtained 
by decimating $n$ atoms, the $\alpha$ and the $\gamma$ sites (colored blue) will have the 
same on site potentials as given earlier, by Eq.~\eqref{rgeps}, viz., 
\begin{equation}
\epsilon_{\alpha (\gamma)} = \epsilon + 2t \frac{U_{n-1}(x)}{U_n(x)}
\end{equation}
Let us now set $E-\epsilon_\alpha = E-\epsilon - 2t U_{n-1}(x)/U_n(x) = 0$, extract the 
(real) roots 
from this equation, and then set $U_m(y)=0$ for one such root. This 
immediately generates a correlation between the values of $\lambda$ and $t$.
 
From Eq,~\eqref{notrap} it is evident that, the third term 
on the right hand side disappears with this value of the tunnel hopping 
$\lambda$, and hence $\epsilon_\beta$ becomes identical 
to $\epsilon_\alpha$ and $\epsilon_\gamma$. We are back with a periodically 
ordered array of identical on-site potentials. 
The stubs effectively become {\it non-existent} to an 
electron propagating in such a system with the above energy.
The chain with the aperiodically placed stub will now transmit the 
electron ballistically at those special energies and {\it no trapping will be observed}.

However, we must appreciate that for each root of the equation 
$E-\epsilon_\alpha=E-\epsilon-2t U_{n-1}(x)/U_n(x)=0$, we shall need 
a different correlation between $\lambda$ and $t$.
For example, for an un-renormalized backbone, that is with $n=0$, if we attach a 
single atom in the side coupled cluster ($m=0$), then the electron with energy $E=0$ 
effectively gets trapped as the lattice grows in size, as explained already. If we 
attach one more site in the side attachment, by making $m=1$, then the effective 
potential at the $\beta$ site becomes same as that of the $\alpha$ and 
the $\gamma$ sites and the electron traverses ballistically at $E=0$. That is, 
for a given relationship between 
$\lambda$ and $t$, just by controlling the number of QD's in the side 
attachment we can make an 
incoming electron get trapped or transmit with undiminished amplitude of its 
wave function. With $m=1$, and with $n=1$, we can make the electron 
tunnel through the entire infinite chain ballistically at $E=0$ if we select 
$\lambda = \pm \sqrt{2}t$. For any other value of $\lambda$, the electron eventually 
gets trapped in quantum wells with increasing height and width as explained before.

\textbf{Controlling trapping energy values by an external magnetic flux.} --
In this part, we show that, it is also possible to `place' the energy eigenvalues
almost anywhere in the spectrum of the ordered backbone by an external magnetic flux.
For this we choose to attach  
quantum rings (QR) to the $\alpha$ and the $\gamma$ sites marked 
(artificially) on the periodic chain of atoms. 
\begin{figure}[ht]
{\centering \resizebox*{7cm}{2.5cm}{\includegraphics{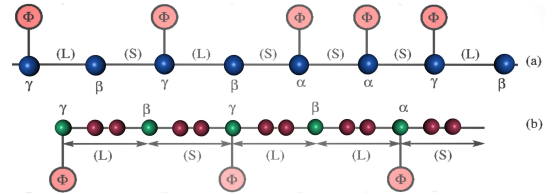}}\par}
\caption{(Color online). (a) Periodically ordered lattice stubbed with a 
quantum ring to generate an {\it artificial} Cantor sequence of vertices and (b) 
periodic repetition of the 3rd generation Fibonacci segment $LSL$, with quantum 
rings side coupled to `create' the Cantor geometry of $\alpha$, $\beta$ and $\gamma$ 
vertices.}
\label{lattice2}
\end{figure}
We refer to Fig.~\ref{lattice2}. The  
quantum rings are are modelled with atomic sites, and are 
attached to the $\alpha$ and $\gamma$ sites on the periodic backbone.
The $\beta$ sites are now free. 
This results in the effective potential at the $\alpha$ and the $\gamma$ 
 sites, given by, 
$\epsilon_\alpha = \epsilon + \chi(E,\Phi)$, where $\chi(E,\phi)$ is the `correction' 
added to the self energy of the bare site, 
obtained after renormalizing the effect of the 
QR. To be specific, for a $4$-site QR threaded by a uniform flux $\Phi$, 
the correction is 
given by, 
\begin{equation}
\chi = \frac{(E-\epsilon) \left [ (E-\epsilon)^2 - 2 t^2 \right ] \lambda^2}
{\left [ (E-\epsilon)^2 - 2 t^2 \right ]^2 - 4t^4 \cos^2\frac{\pi \Phi}{\Phi_0}}
\label{selfen}
\end{equation}
As the trapping energy values are obtained from the roots of the equation
\begin{figure*}[ht]
{\centering \resizebox*{16cm}{5cm}{\includegraphics{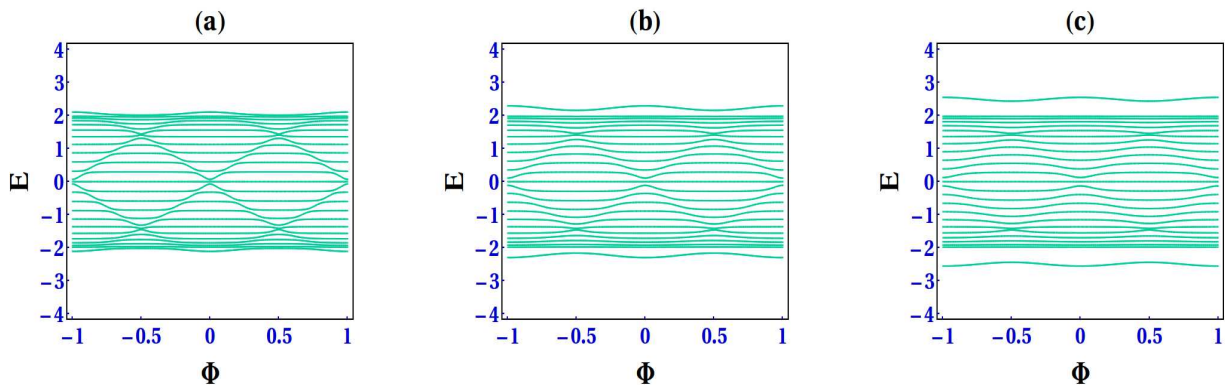}}\par}
\caption{(Color online). Distribution of the trapping energies as the flux through 
a $4$-site quantum ring attached to the $\alpha$ and the $\gamma$ sites on a renormalized 
periodic backbone is varied. We have obtained the renormalized lattice by sequentially 
decimating $33$ atomic site on the initial chain. The tunnel hopping in the three 
panels are chosen an $\lambda=0.5$, $\lambda=1.0$ and $\lambda=2.0$ in units of the 
hopping integral $t$.} 
\label{ordered}
\end{figure*}
$E-\epsilon_\alpha=0$, it is obvious that, the positions of the trapping energies 
in the spectrum can be controlled by tuning the external flux. 
The energy values $E=\epsilon$ and $E=\epsilon\pm \sqrt{2}t$ are excluded 
of course, as at such energies $\chi=0$, and all the sites have identical 
on-site potentials. The propagating electron again doesn't `feel' the stubs, and the 
transport remains ballistic even with the aperiodic stubbing of the backbone. But 
for other energies, the trap is created again.

We present such an example in 
\begin{figure*}[ht]
{\centering \resizebox*{16cm}{5cm}{\includegraphics{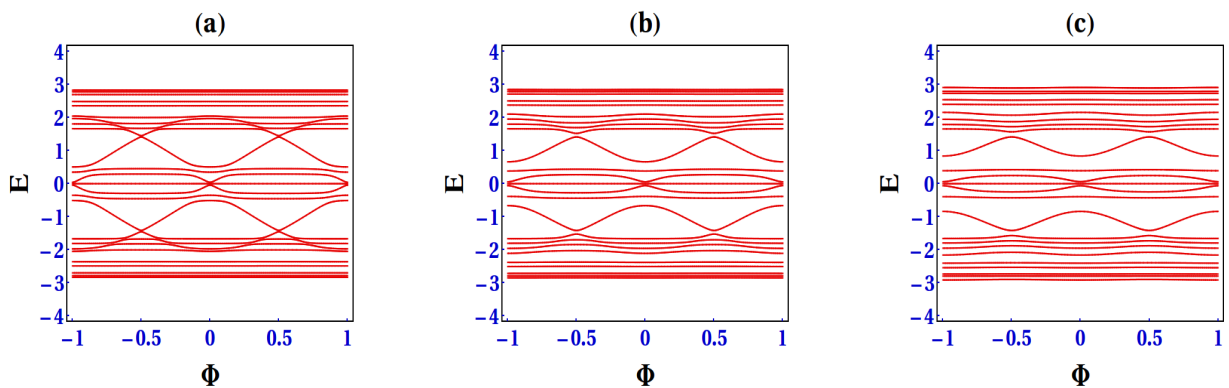}}\par}
\caption{(Color online). Distribution of the trapping energies as the flux through 
a $4$-site quantum ring attached to the $\alpha$ and the $\gamma$ sites on a renormalized 
quasiperiodic Fibonacci backbone is varied. 
We have obtained the renormalized lattice by sequentially 
decimating the vertices on the parent chain by the deflation 
rule $LSL \rightarrow L$, and $LS \rightarrow S$. The tunnel hopping in the three 
panels are chosen an $\lambda=0.5$, $\lambda=1.0$ and $\lambda=2.0$ in units of the 
hopping integral $t$.} 
\label{fibotrap}
\end{figure*}
Fig.~\ref{ordered}, where we first renormalize the periodic backbone 
(no stubs now) by sequentially 
removing $33$ sites uniformly in between two flanking sites in the chain. The renormalized 
periodic chain then has $\epsilon_{eff} = \epsilon + 2t U(32,x)/U(33,x)$, and $t_{eff}=t/U(33,x)$, 
with $x=(E-\epsilon)/2t$. Then the $\alpha$, $\beta$ and $\gamma$ sites are artificially 
marked on this renormalized chain, following the scheme outlined before. 
With the $\alpha$ and 
the $\gamma$ sites we attach the $4$-site QR, and the `corrected' self energies at these sites 
are obtained by adding $\chi(E,\Phi)$, as written in Eq.~\eqref{selfen}. Fig.~\ref{ordered} 
already demonstrates that the trapping energies are distributed over the entire range of the 
allowed spectrum of the periodic backbone, which is between $[-2,2]$ with $\epsilon$ and $t$ 
being chosen as zero and unity respectively. 
The tunnel coupling with the rings gives rise to 
two localized states outside the main band of energies. The localized states get shifted 
in energy as the tunnel hopping strength is increased from $0.5$ to $2.0$ in units of $t$.

\textbf{A fractal distribution of trapping states.} --
It now becomes quite obvious that, the scheme of trapping an electronic excitation 
at specified energies is kind of universal in the sense that, it works irrespective 
of the structure of the `unit cell' as long as we have a periodic lattice of repeating 
unit cells. With `size' of the unit cell increasing, the trapping energies will be 
distributed in the manner dictated by the spectral character of the lattice when the 
relevant unit cell becomes, in principle, infinitely large by itself. To explain this we 
present in Fig.~\ref{fibotrap} the energy eigenvalues controlled by an attached array 
of $4$-site QR's on special sites of a periodic lattice whose unit cell comprises
an 
eighth generation quasiperiodic Fibonacci chain, of two `bonds' $L$ and $S$.

To appreciate, we need to recall a bit that, a quasiperiodic Fibonacci chain is 
a prototype of a quasicrystal in one dimension, consisting of two letters, say $L$ and $S$, 
representing two `bonds' and arranged in a sequence that grows recursively following 
the growth rule $L \rightarrow LS$ and $S \rightarrow L$~\cite{kohmoto}. 
The first few generations 
look like $G_1=L$, $G_2=LS$, $G_3=LSL$, $G_4=LSLLS$ and so on. Any finite generation 
$G_j$ can be periodically repeated along a line to create a periodic chain with 
a quasiperidic geometry in its unit cell, and the spectrum of the true Fibonacci chain 
in its thermodynamic limit is achieved when the size of the unit cell blows up to 
infinity as well. The spectrum, for the sequence outlined above, is singular continuous, 
a Cantor set with measure zero. It typically shows a three subband structure, 
which exhibit self similarity on finer scan, and the wave functions are neither 
{\it extended}, Bloch like, nor are they exponentially localized in the Anderson sense. 
Instead, they are called {\it critical}. 

What we do here is the following. We take, for the sake of explaining things,  
an eighth generation Fibonacci segment 
comprising $34$ bonds $L$ and $S$ set in the Fibonacci sequence, and repeat it 
periodically to generate our periodic chain. Then we scale the system by the standard 
RSRG procedure, using an extension of the usual decimation scheme, viz., $LSL \rightarrow 
L'$ and $LS \rightarrow S'$~\cite{arun}. We need to identify three kinds of on site potentials 
for this. These are, $\epsilon_\alpha$, $\epsilon_\beta$ and $\epsilon_\gamma$ corresponding 
to the vertices flanked by $LL$, $LS$ and $SL$ bonds respectively. The hopping integrals 
are assigned two values, namely, $t_L$ and $t_S$ for an electron hopping across an $L$ 
or an $S$ bond respectively. The decimation of the lattice generates the recursion 
relations, 
\begin{widetext}
\begin{eqnarray}
\epsilon_{\alpha,n} & = & \epsilon_{\alpha,n-1} + \frac{t_{L,n-1}^2 
\left [2E-(\epsilon_{\gamma,n-1}+\epsilon_{\beta,n-1})\right ]}
{(E-\epsilon_{\beta,n-1})(E-\epsilon_{\gamma,n-1})-t_{S,n-1}^2} \nonumber \\
\epsilon_{\beta,n} & = & \epsilon_{\alpha,n-1}+\frac{t_{L,n-1}^2}{(E-\epsilon_{\beta,n-1})}
+\frac{t_{L,n-1}^2 (E-\epsilon_{\beta,n-1})}{(E-\epsilon_{\beta,n-1})
(E-\epsilon_{\gamma,n-1})-t_{S,n-1}^2} \nonumber \\
\epsilon_{\gamma,n} & = & \epsilon_{\gamma,n-1}+ \frac{t_{S,n-1}^2}{E-\epsilon_{\beta,n-1}}
+ \frac{t_{L,n-1}^2 (E-\epsilon_{\beta,n-1})}{(E-\epsilon_{\beta,n-1})
(E-\epsilon_{\gamma,n-1})-t_{S,n-1}^2} \nonumber \\
t_{L,n} & = & \frac{t_{L,n-1}^2t_{S,n-1}}{(E-\epsilon_{\beta,n-1})
(E-\epsilon_{\gamma,n-1})-t_{S,n-1}^2} \nonumber \\
t_{S,n} & = & \frac{t_{L,n-1}t_{S,n-1}}{E-\epsilon_{\beta,n-1}}
\label{fiborecur}
\end{eqnarray}
\end{widetext}
\vskip .5in
\noindent

Using the recursion relations Eq.~\eqref{fiborecur} $2n-1$ times we can `reduce' 
a $2n+1$-th generation Fibonacci segment to a single {\it renormalized} $L$ bond, and 
the periodic lattice now consists of $\alpha$ sites only, with on-site potential 
$\epsilon_{\alpha,2n-1}$ and nearest neighbor hopping integral $t_{L,2n-1}$. On this 
periodic lattice, we {\it artificially} create a Cantor sequence in the spirit described 
before, and attach a $4$-site QR threaded by a magnetic flux at the {\it artificial} 
$\alpha$ and $\gamma$ site to ensure trapping at $E=\epsilon_{\alpha,2n-1}+\chi(E,\Phi)$.

In Fig.~\ref{fibotrap} we show the results, where we have reduced a $34$-bond long 
Fibonacci segment of $L$ and $S$ into an effective periodic lattice and then stubbed 
it with QR's at appropriate sites. 
The initial parameters are chosen as $\epsilon_\alpha=\epsilon_\beta=\epsilon_\gamma=0$, a
and $t_L=1$ and $t_S=2$.
The diagram displays the trapping energy eigenvalues 
at various values of the treading magnetic flux. Interestingly, we find that, even 
with a nominal size of the unit cell, the three-subband structure in the energy 
spectrum, a typical characteristic of a Fibonacci lattice, is quite apparent.
However, now one can see trapping energies located in the global gaps of the 
spectrum. A {\it local Hofstadter butterfly} tends to open up in the central 
part, around $\Phi=0$, as the strength of the tunnel hopping $\lambda$ is increased 
from $0.5$ to $1.0$ and then $2$, in units of $t_L$.

\textbf{Concluding remarks} --

We have shown that a perfectly periodic chain of atomic scatterers can effectively
trap an electron if stubbed at an infinite deterministic set of sites. The trapping, 
which makes an electronic wave function form standing waves in an infinite assembly 
of quantum wells with a hierarchically distributed widths,  
can be effected at any desired scale of length. The energy values at which this takes place 
can be obtained using a real space renormalization group technique.
The scale of length at which the trapping is observed and the locations of the stubs 
are intimately connected, as discussed in the paper. The energy at which one desires 
to trap an electron can be controlled at will by stubbing the lattice with a quantum 
ring threaded by a magnetic field, at a special subset of sites.The method is not 
restricted to the case of a non-interacting spinless electron as stated in this article, but can be 
applied to the propagation of waves, light waves for example, in an array of single mode waveguides.
Work is in progress in this direction and the results will be published in due course.
\vskip .5cm
\vskip 0.15cm
\noindent

\acknowledgments
One of the authors A. Mukherjee is thankful to DST, India for the financial support
provided through research fellowship [Award letter No.: $IF 160437$]. Partial financial support from
University of Kalyani and DST, India through DST-PURSE grant is thankfully acknowledged.



\begin{thebibliography}{99}


\bibitem{anderson} P. W. Anderson, Phys. Rev. \textbf{109}, 1492 (1958).
 
\bibitem{abrahams} E. Abrahams, P. W. Anderson, D. C. Licciardello, and 
T. V. Ramakrishnan, Phys. Rev. Lett. \textbf{42}, 673 (1979).
 
\bibitem{kramer} B. Kramer, and  A. MacKinnon, Rep. Prog. Phys. \textbf{56}, 
1469 (1993).
  
\bibitem{monthus} C. Monthus, and T. Garel, Phys. Rev. B \textbf{81}, 224208 (2010).
 
\bibitem{lyo} S. K. Lyo, Phys. Rev. Lett. \textbf{28}, 1192 (1972).
  
\bibitem{john} S. John, Phys. Rev. Lett. \textbf{58}, 2486 (1987).
  
\bibitem{yablono} E. Yablonovitch, Phys. Rev. Lett. \textbf{58}, 2059 (1987).
  
\bibitem{roati} G. Roati, C. D\'{E}rrico, L. Fallani, M. Fattori, C. Fort, M. Zaccanti, 
G. Modugno, M. Modugno, and M. Inguscio, Nature (London)\textbf{453}, 895 (2008).
 
\bibitem{orellana} P. A. Orellana, F. Dom\'{i}nguez-Adame, I. G\'{o}mez, and M. L. Ladron 
de Guevara, Phys. Rev. B \textbf {67}, 085321 (2003).
  
\bibitem{grosso} R. Farchioni, G. Grosso, and G. P. Parravicini, Phys. Rev. B \textbf{85}, 
165115 (2012).

\bibitem{mirosh} A. E. Miroshnichenko, and Y. S. Kivshar, Phys. Rev. E \textbf{72}, 
056611 (2005).

\bibitem{sergej} A. E. Miroshnichenko, S. Flach, and Y. S. Kivshar, Rev. Mod. Phys. \textbf{82}, 
2257 (2010).
	
\bibitem{dunlap} D. H. Dunlap, K. Kundu, and P. Phillips, Phys. Rev. B \textbf{40}, 
10999 (1989).
	
\bibitem{ac1} A. Chakrabarti, S. N. Karmakar, and R. K. Moitra, Phys. Rev. Lett. \textbf{74}, 1403 (1995).
	
\bibitem{ac2} A. Chakrabarti, S. N. Karmakar, and R. K. Moitra, Phys. Rev. B \textbf{50}, 13276 (1994).

\bibitem{atanu} A. Nandy, B. Pal, and A. Chakrabarti, Europhys. Lett. \textbf{115}, 37004 (2016).
	
\bibitem{biplab1} B. Pal, and A. Chakrabarti, Phys. Lett. A \textbf{378}, 2782 (2014).
	
\bibitem{biplab2} B. Pal, S. K. Maiti, and A. Chakrabarti, Europhys. Lett. \textbf{102}, 17004 (2013).
	
\bibitem{sheelan} S. Sengupta, S. Chattopadhyay, and A. Chakrabarti \textbf{344}, 307 (2004).
	
\bibitem{lind} B. Lindquist, and R. Riklund, Phys. Rev. B \textbf{56}, 13902 (1997).

\bibitem{kohmoto} M. Kohmoto, B. Sutherland, and C. Tang, Phys. Rev. B \textbf{35}, 1020 (1987).
	
\bibitem{arun} A. Chakrabarti , Phys. Rev. B \textbf{74}, 205313 (2006).
	

	
	
  
  
  
  
  

\end{thebibliography}
\end{document}